# Properties on n-dimensional convolution for image deconvolution

Song yizhi[1,2], Xu cheng[1,2], Ding Daoxin[1,2,3,], Zhou Hang[1,2], Quan Tingwei[1,2,3], Li Shiwei[1,2,*]

[1] Collaborative Innovation Center for Biomedical Engineering, Wuhan National Laboratory for Optoelectronics-Huazhong University of Science and Technology, Wuhan, Hubei 430074, China

[2] Britton Chance Center and MOE Key Laboratory for Biomedical Photonics, School of Engineering Sciences, Huazhong University of Science and Technology, Wuhan, Hubei 430074, China

[3] School of Mathematics and Economics, Hubei University of Education, Wuhan, Hubei 430205, China

Email: swingsw89@163.com

## Abstract

Convolution system is linear and time invariant, and can describe the optical imaging process. Based on convolution system, many deconvolution techniques have been developed for optical image analysis, such as boosting the space resolution of optical images, image denoising, image enhancement and so on. Here, we gave properties on N-dimensional convolution. By using these properties, we proposed image deconvolution method. This method uses a series of convolution operations to deconvolute image. We demonstrated that the method has the similar deconvolution results to the state-of-art method. The core calculation of the proposed method is image convolution, and thus our method can easily be integrated into GPU mode for large-scale image deconvolution.

# Introduction

Optical imaging techniques broaden our views of objects that cannot be observed directly through our eyes [1-2], and have the widespread applications in many fields [3-6] such as biological research and medicine research. Though there exist a great variety of developed optical imaging techniques, almost all of them are based on a convolution system [7-9]. In this convolution system, the observed image is generated by the target image convoluted with a point spread function. This convolution process can also be described by a group of linear equations because image convolution is a linear transformation [9]. By solving the linear equations, the target image can be generated from observed image and the point spread function. This inverse process is called image deconvolution. In general, compared to the observed image, the target image has higher spatial resolution and signal-to-noise ratio [9-10]. So, image deconvolution has been one of the most important topics in imaging techniques.

Many efforts have been made for developing image deconvolution methods. In early years, constrained iterative deconvolution [11] was used for wide-field microscopy. Aiming at confocal images, some typical methods have been proposed including Tikhonov–Miller inverse filter [12], Carrington [13] and Richardson–Lucy (RL) algorithms [14-16]. These methods introduce the regularization constraints to avoid the sensitivity to noise, and vastly boost the results. Recently, many novel image techniques inspire image deconvolution. For example, Total variation (TV) regularization [18-19] has shown its widespread applications such as image denoising, inpainting, segmentation and so on. TV regularization naturally extends to image deconvolution. In addition, sparse reconstruction techniques also can be used into image deconvolution [20-22].

In fact, most of image deconvolution methods introduce prior knowledge, expressed by regularization term [23-28], for robust target image. When regularization term is added into the framework of image deconvolution, image deconvolution is usually taken to solve the designed optimization problem. Generating optimal solutions requires to solve a number of linear equations, and FFTs are repeatedly employed for this purpose [9]. We gave properties on convolution system, which leads to another way to generate optimal solution. Namely, image deconvolution can be simply achieved through a series of image convolution operations. We verified the effectiveness of our method. This kind of operation can be combined into GPU mode and thus has high potential for large-scale image deconvolution.

## Properties on n-dimensional convolution

As described above, convolution system is linear and can be described by a group of linear equations. The corresponding coefficient matrix has the specific structure. We presented this structure driven from n-dimensional convolution system. According to this structure, we built the links between convolution operation and the transpose of the corresponding coefficient matrix. We summarized the above characteristics of convolution system and gave the corresponding note by using the two following properties.

**Property 1**. For two n-dimensional discrete functions, respectively, denoted by

$$x(\mathrm{s}_1,...,s_n),\quad (\mathrm{s}_1,...,s_n)\in A=\{s_1=0,1,\cdots,m_1;\cdots;s_n=0,1,\cdots,m_n\}\ \text{ and}$$

$$h(\mathrm{s}_1,...,s_n),\quad (\mathrm{s}_1,...,s_n)\in B=\{s_1=0,1,\cdots,2p_1;\cdots;s_n=0,1,\cdots,2p_n\}.$$

Their discrete convolution is defined as,

$$y(\mathrm{s}_1,...,\mathrm{s}_n)=\sum_{k_1=1}^{m_1}\sum_{k_2=1}^{m_2}...\sum_{k_n=1}^{m_n}x(k_1,...,k_n)\,\mathrm{h}(\mathrm{s}_1-k_1,...,s_n-k_n)$$

Rewrite the above functions into matrix, then the discrete convolution is

$$Y=X\otimes\mathrm{H}$$

Where, Y, X and H are n-dimensional matrix, denoted by $Y[y_{\mathrm{s}_1,...,\mathrm{s}_n}]$ $X[x_{\mathrm{s}_1,...,s_n}]$ and $\mathrm{H}[h_{s_1,...,s_n}]$, respectively.

The matrix $Y$ and $X$ can be converted into two column vectors, which are $Y_n^v$ and $X_n^v$ respectively, we have

$$Y_n^v=\mathrm{A}_n^hX_n^v \tag{1}$$

$$\mathrm{Y}_n^v = \begin{bmatrix} Y_n^v(0,\cdots) \\ Y_n^v(1,\cdots) \\ \vdots \\ \vdots \\ Y_n^v(m_1+2p_1,\cdots) \end{bmatrix}, \quad X_n^v = \begin{bmatrix} X_n^v(0,\cdots) \\ X_n^v(1,\cdots) \\ \vdots \\ \vdots \\ X_n^v(m_1,\cdots) \end{bmatrix},$$

$$A_n^h = \begin{bmatrix} A_n^h(0,\cdots) & 0 & \cdots & 0 \\ A_n^h(1,\cdots) & A_n^h(0,\cdots) & \cdots & 0 \\ \vdots & \vdots & \ddots & \vdots \\ A_n^h(2p_1,\cdots) & A_n^h(2p_1-1,\cdots) & & \vdots \\ 0 & A_n^h(2p_1,\cdots) & & A_n^h(0,\cdots) \\ \vdots & \vdots & & \vdots \\ 0 & 0 & \cdots & A_n^h(2p_1,\cdots) \end{bmatrix}.$$

Here $Y_n^v(j,\cdots),\ j=0,1,\cdots,m_1+2p_1$, and $\mathrm{X}_n^v(j,\cdots),\ j=0,1,\cdots,m_1$ are column vectors driven from (n-1)-dimensional matrix $Y[y_{j,s_2,\ldots,s_n}]$ and $X[x_{j,s_2,\ldots,s_n}]$,respectively, by partitioning along rows.

$A_{n-1}^h(j,\cdots),\ j=0,1,\cdots,2p_1$ is matrix driven from convolution template $\mathrm{H}[h_{j,s_2,\ldots,sn}]$.

The proof provided in Supplementary.

**Property 2**. For a n-dimensional discrete convolution

$$y(\mathrm{s}_1,\ldots,\mathrm{s}_n) = \sum_{k_1=1}^{m_1}\sum_{k_2=1}^{m_2}\ldots\sum_{k_n=1}^{m_n} x(k_1,\ldots,k_n)\,\mathrm{h}(\mathrm{s}_1-k_1,\ldots,s_n-k_n)$$

Its corresponding matrix expression is

$$Y = X \otimes \mathrm{H}.$$

We have

$$(\mathrm{A}_n^h)^T Y^v = \left\{M_{2p_1,\cdots,2p_n}(Y\otimes\tilde{H})\right\}^v = \left\{M_{2p_1,\cdots,2p_n}(\widehat{Y}(\widehat{y}_{s_1,s_2,\cdots,s_n}))\right\}^v \tag{2}$$

Where, $\tilde{H}=H(h_{2p_1-s_1,\cdots,2p_n-s_n})$ and matrix $\widehat{Y}(\widehat{y}_{s_1,s_2,\cdots,s_n})$ whose size is

$$(m_1+4p_1+1)\times(m_2+4p_2+1)\times\cdots\times(m_n+4p_n+1).$$

We defined $M_{2p_1,\cdots,2p_n}$ as $M_{2p_1,\cdots,2p_n}(\widehat{Y}(\widehat{y}_{s_1,s_2,\cdots,s_n})=\widehat{Y}^c(\widehat{y}_{\widehat{s}_1,\widehat{s}_2,\cdots,\widehat{s}_n})$, where

$\widehat{s}_1 = 2p_1+1, 2p_1+2,\cdots,m_1+2p_1+1$, $\widehat{s}_2 = 2p_2+1, 2p_2+2,\cdots,m_2+2p_2+1$, …,

$\widehat{s}_n = 2p_n+1, 2p_n+2,\cdots,m_n+2p_n+1$.

## Application to 2-dimensional deconvolution

Let *X(x, y)* and *Y(x, y)* be input and observed image respectively. The relation between these two images is modeled as

$$Y(x,y) = X(x,y) \otimes H(x,y) + N(x,y) \tag{3}$$

Here, *H(x, y)* is the point spread function, and *N(x, y)* represents noise. The discretization of E.Q (3) is given by

$$Y(y_{s_1,s_2})_{(m+2p_1)\times(n+2p_2)} = X(x_{s_1,s_2})_{m\times n} \otimes H(h_{s_1,s_2})_{(2p_1+1)\times(2p_2+1)} + N(n_{s_1,s_2})_{(m+2p_1)\times(n+2p_2)} \tag{4}$$

According to *Property 1*, E.Q (4) can be described by a group of linear equation, see below.

$$Y_2^v = A_2^h X_2^v + N_2^v \tag{5}$$

Image deconvolution is essential for solving E.Q (5) in which $Y_2^v$ and $A_2^h$ are known. Considering nonnegative solver of $X_2^v$, corresponding to the nonnegative signal intensity of input image X, Solving E.Q (5) is usually converted into minimizing the following object function.

$$\min_{X_2^v \in \mathbb{R}^{mn}} f(X_2^v) = \frac{1}{2}\left\| A_2^h X_2^v - Y_2^v \right\|_2^2$$
$$s.t. X_2^v \geq 0$$

Gradient method is commonly used to solve optimization problem. The corresponding algorithm is described below.

*Step 1*) Set initial solution $\left(X_2^v\right)^0$

*Step 2*) $\left(X_2^v\right)^{k+1} = \max\left\{\left(X_2^v\right)^k - \delta_k\left(A_2^{h\,T} A_2^h (X_2^v)^k - A_2^{h\,T} Y_2^v\right), 0\right\}$

*Step 3*) Repeat *Step 2*) until the solution converges.

Note that, in step 2), the parameter $\delta_k$ is confirmed by the rule, $f((X_2^v)^{k+1}) < f((X_2^v)^k)$.

According to *Properties* 1 and 2, we have

$$
\begin{aligned}
&A_2^{h\,T} A_2^h (X_2^v)^k - A_2^{h\,T} Y_2^v \\
&= \left\{ \mathrm{M}_{2p_1,2p_2} (H(2p_1 - s_1, 2p_2 - s_2) \otimes (H(s_1,s_2) \otimes X(s_1,s_2))) \right\}^v \\
&\quad - \left\{ \mathrm{M}_{2p_1,2p_2} (H(2p_1 - s_1, 2p_2 - s_2) \otimes Y(s_1,s_2))) \right\}^v
\end{aligned} \tag{6}
$$

So, E.Q (6) indicates that solving the optimization problem, namely image deconvolution, can be achieved by performing a series of convolution operations.

## Results

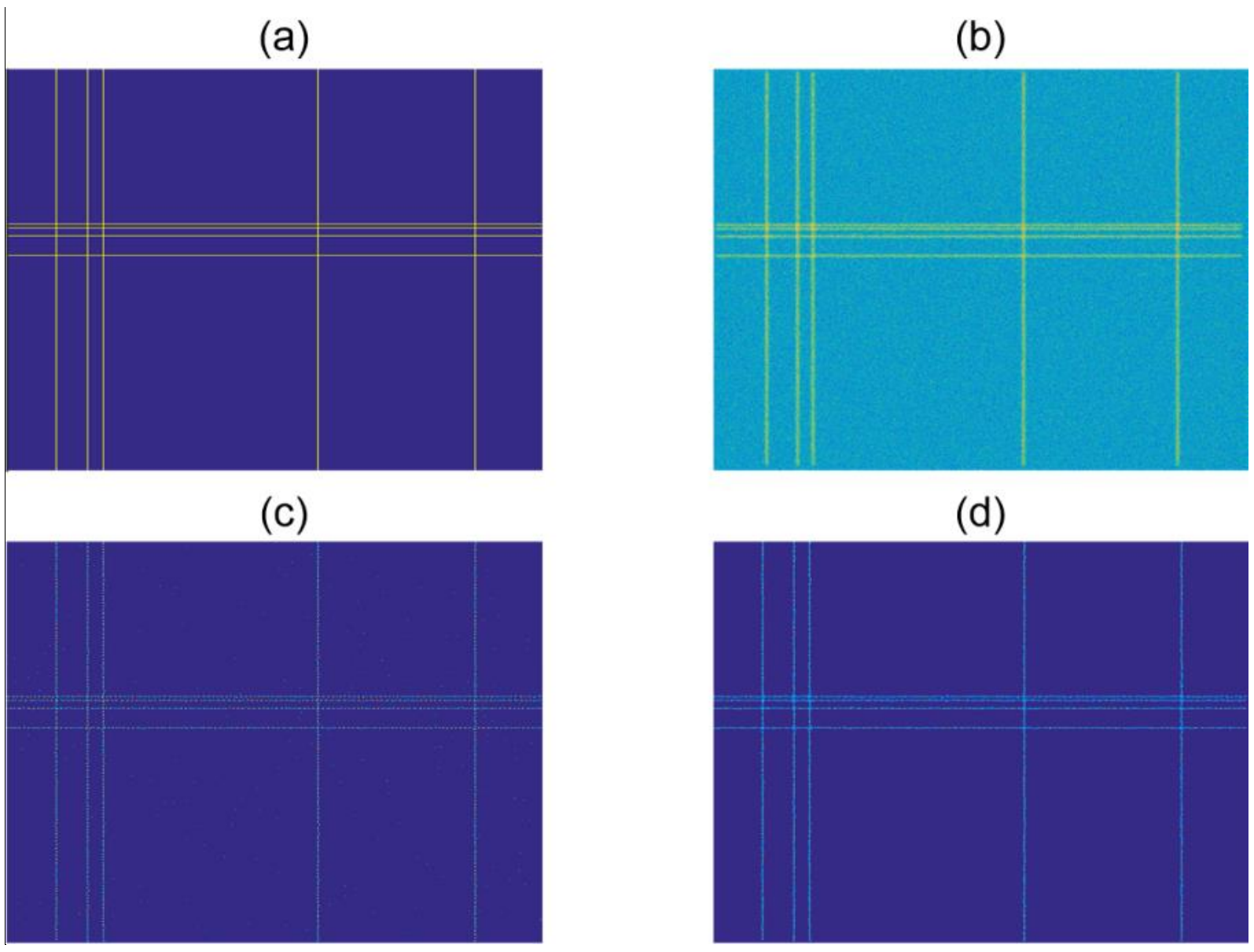

**Figure 1**: Deconvolute the Simulated image with two methods. (**a**) Original image; (**b**) The blurred image by convolving image in (**a**) with Gaussian white noise (mean: 0; standard deviation: 5); Image deconvoluted with our method (**c**) and Richardson–Lucy method (**d**) respectively.

We first use simulated images to confirm that our method can acquire deconvolution results on blurred image.

The target image (**Fig. 1a**) has the size of 512 x 512 pixels, including nine lines distributed crossingly. The target image was convoluted with Gaussian white noise (mean: 0, standard deviation: 5). After this operation, the observed image was obtained (**Fig. 1b**). Using our method and R-L method to deconvole the image in (**Fig. 1b**), the results are presented in (**Fig. 1c**) and (**Fig.1d**) respectively. The deconvoluting results indicate that our deconvolution method performed a good recovery capacity as R-L method did. We conclude that our method can achieve image deconvolution through a series of convolution operation and share similar recovery result when compared to the state of the are method.

We furthermore demonstrated that our method has strong blurring ability in a more complex condition. We used a classic image (8 bit, 255 for maximum intensity value) for this experiment (**Fig. 2a**). The size of this image is $225 \times 225$ pixels. A $5 \times 5$ pixels convolution template is applied to simulate the PSF of lens. The original image was convoluted with Gaussian white noise (mean: 0) and in different standard deviation, respectively. When the noise level is high, (standard deviation: 20), our method can deblur the image with a good anti-noise property (**Fig. 2b**), like R-L method (**Fig. 2c**). When standard deviation of noise is low (standard deviation: 5) (**Fig. 2d**), deconvolution method (**Fig. 2e**) performed as well as R-L method did (**Fig. 2f**). Both deconvoluted images are very similar to the original image (**Fig. 2a**). The above results indicate that method is equipped with high anti-noise property and can deal with strong noisy images.

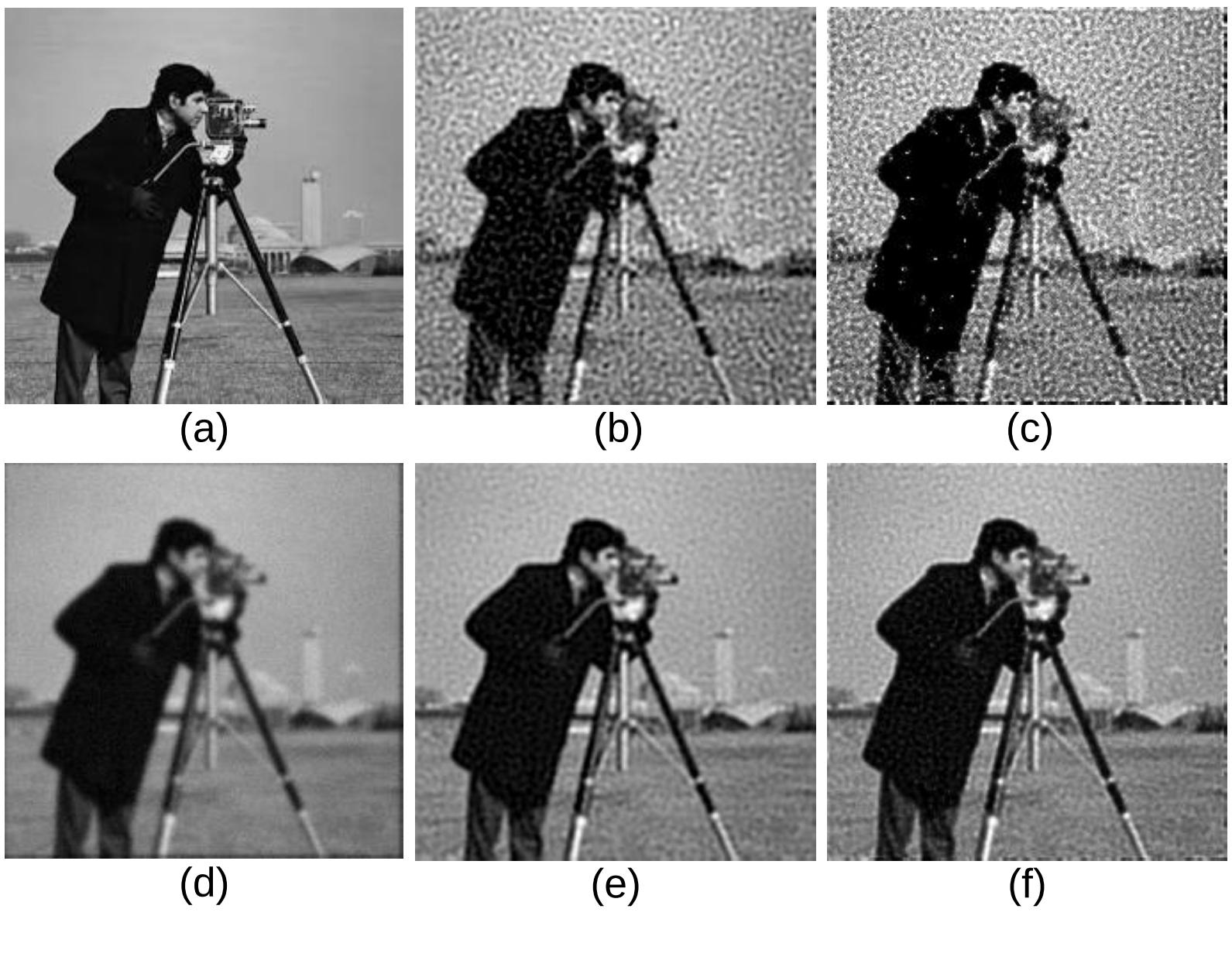

**Figure 2**: comparisons of deconvolution result from multiple methods on images with different noisy level(a)Original image (b) deconvolution result derived by our method with high-noise level (Gaussian white noise (mean: 0, standard deviation: 20)(c) deconvolution result derived by maximum likelihood method  others are the same with (b)(d) a blurred image with Gaussian white noise (mean: 0, standard deviation: 5) (e) deconvolution result derived by our method with high-noise level (Gaussian white

noise (mean: 0, standard deviation: 5), SNR= 19.606 (f) by maximum likelihood method, others are the same with (e)SNR=18.775.

## Conclusions

Image deconvolution has wide applications in various of image processing techniques. In this paper, we first analyze the connections between convolution operation and the transpose of the corresponding coefficient matrix. Then, we verified the image deconvolution operation can be fulfilled through a series of image convolution operations. We confirmed this new method shall similar deconvolution result when compared to the state-of-the art methods. In addition, we provide a vivid prove of our deconvolution method.

## Supplementary

### *Proof of Property 1*

**Proof**：Proved with mathematical induction. When $n$=1, we have

$y(s_1)=\sum_{k_1=0}^{m_1+2p_1+1} x(k_1)h(s_1-k_1), \qquad (k_1=1,2,...,m_1+2p_1+1)$ , and easily obtain

$$\begin{bmatrix} y(0) \\ y(1) \\ \vdots \\ y(m_1+2p_1) \end{bmatrix} = \begin{bmatrix} h(0) & 0 & & 0 \\ h(1) & h(0) & \cdots & 0 \\ \vdots & \vdots & \cdots & \vdots \\ h(2p_1) & h(2p_1-1) & \ddots & \vdots \\ 0 & h(2p_1) & & h(0) \\ \vdots & \vdots & \cdots & \vdots \\ 0 & 0 & & h(2p_1) \end{bmatrix} \begin{bmatrix} x(0) \\ x(1) \\ \vdots \\ x(m_1) \end{bmatrix}. \qquad (S1)$$

Namely, $Y_1^v = A_1^h X_1^v$ holds.

Assume that for $n$, we have $Y_n^v = A_n^h X_n^v$, we prove that $Y_{n+1}^v = A_{n+1}^h X_{n+1}^v$ holds. According to the definition of discrete convolution, we have

$$y(s_1,...,s_n,s_{n+1}) = \sum_{k_1=0}^{m_1}\sum_{k_2=0}^{m_2}\cdots\sum_{k_{n+1}=0}^{m_{n+1}} x(k_1,...,k_n,k_{n+1})h(s_1-k_1,...,s_n-k_n,s_{n+1}\text{-}k_{n+1}) \qquad (S2)$$

When fixing the pair $(s_1,k_1)$, the following formulation can be generated.

$$f(s_1,s_1-k_1) = \sum_{k_2=0}^{m_2}\sum_{k_3=0}^{m_3}\cdots\sum_{k_{n+1}=0}^{m_{n+1}} x(k_1,...,k_n,k_{n+1})h(s_1-k_1,...,s_n-k_n,s_{n+1}\text{-}k_{n+1}) \qquad (S3)$$

This is n-dimensional discrete convolution. Based on the assumption, we obtain

$$F_n^v(s_1,s_1-k_1) = A_{n+1}^h(s_1-k_1,\cdots)X_{n+1}^v(k_1,\cdots) \qquad (S4)$$

According to (S4), we have

$$Y_{n+1}^v(s_1,\cdots) = \sum_{k_1=0}^{m_1} F_n^v(s_1,s_1-k_1) = \sum_{k_1=0}^{M_1} A_{n+1}^h(s_1-k_1,\cdots)X_{n+1}^v(k_1,\cdots), s_1=0,1,\cdots,m_1+2p_1 \qquad (S5)$$

Combine the $m_1$+$2p_1$ groups of linear equations in (S5) into one group of linear equations, and obtain $Y_{n+1}^v = A_{n+1}^h X_{n+1}^v$. The proof is completed.

*Proof of Property 2*

Proof：For n=1, we prove that the conclusion holds. According to *Property* 1, we have

$$A_h^1=\begin{bmatrix} h(0) & 0 & & 0 \\ h(1) & h(0) & \cdots & 0 \\ \vdots & \vdots & \cdots & \vdots \\ h(2p_1) & h(2p_1-1) & \ddots & \vdots \\ 0 & h(2p_1) & \cdots & h(0) \\ \vdots & \vdots & & \vdots \\ 0 & 0 & & h(2p_1) \end{bmatrix}. \tag{S7}$$

The transpose of $A_h^1$ given by

$$\left(A_h^1\right)^T=\begin{bmatrix} h(0) & h(1) & \cdots & h(2p) & \cdots & 0 \\ 0 & h(0) & \cdots & h(2p-1) & \cdots & 0 \\ \vdots & \vdots & \vdots & \vdots & \vdots & \vdots \\ 0 & 0 & \cdots & 0 & \cdots & h(2p) \end{bmatrix}. \tag{S8}$$

We define the extended $\left(A_h^1\right)^T$ as

$$(\widehat{A}_1^h)^T=\begin{bmatrix} h(2p) & 0 & \cdots & 0 & \cdots & 0 \\ \vdots & \vdots & \vdots & 0 & \cdots & 0 \\ h(0) & h(1) & \cdots & h(2p) & \cdots & 0 \\ 0 & h(0) & \cdots & h(2p-1) & \cdots & 0 \\ \vdots & \vdots & \vdots & \vdots & \vdots & \vdots \\ 0 & 0 & \cdots & 0 & \cdots & h(2p) \\ \vdots & \vdots & \vdots & \vdots & \cdots & \\ 0 & 0 & & 0 & 0 & h(0) \end{bmatrix}. \tag{S9}$$

According to *Property* 1， we have

$$(\widehat{A}_1^h)^T Y_1^v=\{Y(y_{s_1})\otimes H(h_{2p_1-s_1})\}^v$$

Based on relation between $\left(A_h^1\right)^T$ and $(\widehat{A}_1^h)^T$， we obtain

$$\left(A_h^1\right)^T Y_1^v=\{M_{2p_1}(Y(y_{s_1})\otimes H(h_{2p_1-s_1}))\}^v.$$

Assume that the conclusion holds for $n$. Now, we prove that the conclusion holds for $n+1$.

We give the matrix driven from the convolution template and its corresponding extended matrix, denoted by

$$A_n^h = \begin{bmatrix} A_n^h(0,\cdots) & 0 & \cdots & 0 \\ A_n^h(1,\cdots) & A_n^h(0,\cdots) & \cdots & 0 \\ \vdots & \vdots & \ddots & \vdots \\ A_n^h(2p_1,\cdots) & A_n^h(2p_1-1,\cdots) & & \vdots \\ 0 & A_n^h(2p_1,\cdots) & & A_n^h(0,\cdots) \\ \vdots & \vdots & & \vdots \\ 0 & 0 & \cdots & A_n^h(2p_1,\cdots) \end{bmatrix}, \tag{S10}$$

$$(\widehat{A}_{n+1}^h)^T = \begin{bmatrix} (\widehat{A}_{n+1}^h(2p_1,\cdots))^T & 0 & \cdots & 0 & \cdots & 0 \\ \vdots & \vdots & \vdots & 0 & \cdots & 0 \\ (\widehat{A}_{n+1}^h(0,\cdots))^T & (\widehat{A}_{n+1}^h(1,\cdots))^T & \cdots & (\widehat{A}_{n+1}^h(2p_1,\cdots))^T & \cdots & 0 \\ 0 & (\widehat{A}_{n+1}^h(0,\cdots))^T & \cdots & (\widehat{A}_{n+1}^h(2p_1-1,\cdots))^T & \cdots & 0 \\ \vdots & \vdots & \vdots & \vdots & \vdots & \vdots \\ 0 & 0 & \cdots & 0 & \cdots & (\widehat{A}_{n+1}^h(2p_1,\cdots))^T \\ \vdots & \vdots & \vdots & \vdots & \cdots & \\ 0 & 0 & & 0 & 0 & (\widehat{A}_{n+1}^h(0,\cdots))^T \end{bmatrix} \tag{S11}$$

According to the assumption that it holds for $n$ and *property* 1, we have

$$\begin{aligned} &(\widehat{A}_{n+1}^h(i,\cdots))^T Y_{n+1}^v(j,\cdots) \\ &= \left\{H(h_{2p_1-i,2p_2-s_2,\cdots,2p_n-s_n}) \otimes Y(y_{j,s_2,\cdots,s_n})\right\}^v \end{aligned} \tag{S12}$$

$$\begin{aligned} &(A_{n+1}^h(i,\cdots))^T Y_{n+1}^v(j,\cdots) \\ &= \left\{\mathrm{M}_{2p_2,\cdots,2p_{n+1}}\{H(h_{2p_1-i,2p_2-s_2,\cdots,2p_n-s_n}) \otimes Y(y_{j,s_2,\cdots,s_n})\}\right\}^v \end{aligned} \tag{S13}$$

$$\begin{aligned} (\widehat{A}_{n+1}^h)^T Y_{\mathrm{n}+1}^v &= \left\{Y(y_{s_1,\ldots,s_n,s_{n+1}}) \otimes \mathrm{H}(h_{2p_1-s_1,\ldots,2p_n-s_n,2p_{n+1}-s_{n+1}})\right\}^v \\ &= \begin{pmatrix} \sum_{k_1=0}^{m_1} \widehat{A}_{n+1}^h(0-k_1,\cdots)\mathrm{Y}_{n+1}^v(k_1,\cdots) \\ \sum_{k_1=0}^{m_1} \widehat{A}_{n+1}^h(1-k_1,\cdots)\mathrm{Y}_{n+1}^v(k_1,\cdots) \\ \vdots \\ \sum_{k_1=0}^{m_1} \widehat{A}_{n+1}^h(m_1+2p_1-k_1,\cdots)\mathrm{Y}_{n+1}^v(k_1,\cdots) \end{pmatrix} \end{aligned} \tag{S14}$$

$$\left\{\mathrm{M}_{0,2p_2,2p_3,2p_{n+1}}\left\{Y(y_{s_1,\ldots,s_n,s_{n+1}})\otimes\mathrm{H}(h_{2p_1-s_1,\ldots,2p_n-s_n,2p_{n+1}-s_{n+1}})\right\}\right\}^{v}$$

$$=\begin{pmatrix} \sum_{k_1=0}^{m_1} A_{n+1}^{h\ T}(0-k_1,\cdots)\mathrm{Y}_{n+1}^{v}(k_1,\cdots) \\ \sum_{k_1=0}^{m_1} A_{n+1}^{h\ T}(1-k_1,\cdots)\mathrm{Y}_{n+1}^{v}(k_1,\cdots) \\ \vdots \\ \sum_{k_1=0}^{m_1} A_{n+1}^{h\ T}(m_1+2p_1-k_1,\cdots)\mathrm{Y}_{n+1}^{v}(k_1,\cdots) \end{pmatrix} \tag{S15}$$

According to S12-S15, we have

$$A_{n+1}^{h\ T}Y_{n+1}^{v}=\left\{\mathrm{M}_{2p_1,2p_2,2p_3,2p_{n+1}}\left\{Y(y_{s_1,\ldots,s_n,s_{n+1}})\otimes\mathrm{H}(h_{2p_1-s_1,\ldots,2p_n-s_n,2p_{n+1}-s_{n+1}})\right\}\right\}^{v}.$$

The proof is completed.